\documentclass[12pt,preprint]{aastex}

\def\kms{\ifmmode{\rm km\thinspace s^{-1}}\else km\thinspace s$^{-1}$\fi}

\shortauthors{Torres}
\shorttitle{Orbit for HD~34700}

\begin{document}

\title{A double-lined spectroscopic orbit for the young star HD~34700}

\author{Guillermo Torres}

\affil{Harvard-Smithsonian Center for Astrophysics, 60
Garden St., Mail Stop 20, Cambridge, MA 02138}

\email{gtorres@cfa.harvard.edu}

\begin{abstract} 

We report high-resolution spectroscopic observations of the young star
HD~34700, which confirm it to be a double-lined spectroscopic binary.
We derive an accurate orbital solution with a period of $23.4877 \pm
0.0013$~days and an eccentricity of $e = 0.2501 \pm 0.0068$. The stars
are found to be of similar mass ($M_2/M_1 = 0.987 \pm 0.014$) and
luminosity. We derive also the effective temperatures (5900~K and
5800~K) and projected rotational velocities (28~\kms\ and 22~\kms) of
the components. These values of $v \sin i$ are much higher than
expected for main-sequence stars of similar spectral type (G0), and
are not due to tidal synchronization. We discuss also the indicators
of youth available for the object. Although there is considerable
evidence that the system is young ---strong infrared excess, X-ray
emission, \ion{Li}{1}~$\lambda$6708 absorption (0.17~\AA\ equivalent
width), H$\alpha$ emission ($\sim$0.6~\AA), rapid rotation--- the
precise age cannot yet be established because the distance is unknown. 

\end{abstract}

\keywords{binaries: spectroscopic --- stars: evolution --- stars:
individual (HD~34700) --- stars: pre--main-sequence --- techniques:
spectroscopic}

\section{Introduction}
\label{sec:intro}

Over the past decade numerous surveys of young stars in star-forming
regions have found evidence that the overall frequency of binary and
multiple systems is consistently higher than in the general field
\citep[e.g.,][]{Ghez:93,Leinert:93,Ghez:97a,Ghez:97b,Kohler:98,Simon:99a},
although the enhancement may vary somewhat from region to region
\citep[see, e.g.,][]{Simon:99b}. In general those surveys have
indicated roughly twice as many binaries as in the classical study by
\cite{Duquennoy:91}, which focussed on solar-type stars in the solar
neighborhood that are typically much older. The results for the
younger binaries refer mostly to relatively wide systems ($> 5$~AU),
spatially resolved by high-resolution imaging techniques such as
infrared speckle interferometry or adaptive optics \citep[see,
e.g.,][]{Takami:03}.  Studies of binaries with smaller separations
(spectroscopic binaries) in star-forming regions have not been as
systematic, but may indicate a similar trend \citep[see,
e.g.,][]{Mathieu:94,Corporon:99,Zinnecker:01,Melo:03}. 

Partial lists of known spectroscopic binaries among pre--main sequence
(PMS) stars have been presented by \cite{Melo:01}, \cite{Prato:02},
and others, and new binaries are slowly but continuously being added.
More than two dozen of the systems with known orbits are double-lined
spectroscopic binaries, which provide the most information on the
individual masses of the components (compared to single-lined
spectroscopic binaries). 

The young star HD~34700 has recently been reported to be a
double-lined spectroscopic binary by \cite{Arellano:03}, who also
provided evidence of velocity variations. HD~34700 was originally
identified as a young object from its strong infrared excess as
observed by IRAS \citep{Odenwald:86,Walker:88,Oudmaijer:92}, a
characteristic usually interpreted in terms of a circumstellar disk in
other similar ``Vega-excess" stars. It is located in Orion, although
it is not clear whether it is associated with any of the star-forming
complexes in the general vicinity.  The optical, infrared, and
millimeter-wave properties of HD~34700 have been modeled extensively
by \cite{Sylvester:96a} and \cite{Sylvester:96b}.  They inferred that
the disk has inner and outer radii of roughly 25--50~AU and 550~AU,
respectively, depending on the model, and that the circumstellar
material is composed mostly of relatively small dust grains (size
$\leq 10~\mu$m).  Sub-millimeter observations have led to the
detection of $^{12}$CO and $^{13}$CO emission, from which a total mass
of $\sim$1~M$_{\earth}$ in dust particles has been derived
\citep{Zuckerman:95,Coulson:98}.  Other estimates have varied between
0.2~M$_{\earth}$ and 3.7~M$_{\earth}$
\citep{Sylvester:96b,Walker:00,Sylvester:01}, depending also on the
assumed distance.  Molecular hydrogen in the disk has been searched
for, but not seen \citep{Bary:03}. Low-level optical linear
polarization has been detected by \cite{Bhatt:00} and
\cite{Oudmaijer:01}, consistent with a non-spherically symmetric
distribution of dust (a disk).  The infrared luminosity of the star is
a considerable fraction of its bolometric luminosity
\citep[$\sim$8--35\%;][]{Zuckerman:95,Sylvester:96a,Coulson:98,Bhatt:00,Sylvester:01}.
As in other young objects, the \ion{Li}{1}~$\lambda$6708 absorption
line is quite prominent in HD~34700, and the H$\alpha$ line is seen in
emission with highly variable profiles
\citep{Zuckerman:94,Arellano:03}.  It is also a strong X-ray source.
The spectral type of the star has been listed as G0
\citep{Odenwald:86}, or more recently as G\ion{0}{4}e \citep{Mora:01}. 

Given the interest in the object and the evidence that it may be a
spectroscopic binary, the main motivation for this paper is to present
high-resolution spectroscopic observations that indeed confirm its
double-lined nature. From these observations we derive an accurate
double-lined orbital solution. We discuss also the available
indicators of youth. 

\section{Observations}
\label{sec:obs}

HD~34700 (also HIP~24855, SAO~112630, IRAS~05170+0535, $\alpha =
5^{\rm h} 19^{\rm m} 41\fs41$, $\delta = +5\arcdeg 38\arcmin
42\farcs8$, J2000, $V = 9.15$) was originally placed on our observing
program at the Harvard-Smithsonian Center for Astrophysics (CfA) in
1996 as part of a project to monitor the radial velocities of several
dozen nearby stars believed to be young from a variety of
spectroscopic and photometric indicators (infrared excess, H$\alpha$
emission, strong \ion{Li}{1}~$\lambda$6708 absorption, etc.).
Observations were obtained mostly with an echelle spectrograph on the
1.5-m Wyeth reflector at the Oak Ridge Observatory (Harvard,
Massachusetts), and occasionally also with a nearly identical
instrument on the 1.5-m Tillinghast reflector at the F.\ L.\ Whipple
Observatory (Mt.\ Hopkins, Arizona). A single echelle order centered
at 5187~\AA\ was recorded using intensified photon-counting Reticon
detectors, giving a spectral window of 45~\AA. The resolving power of
these observations is $\lambda/\Delta\lambda \approx 35,\!000$. The
signal-to-noise (S/N) ratios range mostly from about 20 to 30 per
resolution element of 8.5~\kms, with the exception of our first
exploratory exposure. 

That observation in March of 1996 clearly revealed the double-lined
nature of the system from the double peaks in the cross-correlation
function, despite being a very weak spectrum with a S/N ratio of only
7. The relative strengths of the peaks were similar, indicating stars
of nearly equal luminosity. Once we realized this, exposure times were
increased accordingly. Subsequent observations confirmed the double
peaks, and we continued to monitor the star for another four years,
obtaining a total of 35 spectra. 

\section{Radial velocities and orbital solution}
\label{sec:orbit}

Radial velocities were derived using TODCOR \citep{Zucker:94}, a
two-dimension\-al cross-correlation algorithm well suited to our
relatively low S/N spectra. TODCOR uses two templates, one for each
component of the binary, and combines one-dimensional correlation
functions into a two-dimensional function that avoids the common
blending problems of the standard procedures. The one-dimensional
correlation functions were computed using the IRAF\footnote{IRAF is
distributed by the National Optical Astronomy Observatories, which is
operated by the Association of Universities for Research in Astronomy,
Inc., under contract with the National Science Foundation.} task XCSAO
\citep{Kurtz:98}. The templates were selected from a large library of
synthetic spectra based on model atmospheres by R.\ L.\
Kurucz\footnote{Available at {\tt http://cfaku5.cfa.harvard.edu}.},
computed for us by Jon Morse \citep[see
also][]{Nordstrom:94,Latham:02}. These calculated spectra are
available for a wide range of effective temperatures ($T_{\rm eff}$),
projected rotational velocities ($v \sin i$), surface gravities ($\log
g$) and metallicities. Experience has shown that radial velocities are
largely insensitive to the surface gravity and metallicity adopted for
the templates. Consequently, the optimum template for each star was
determined from grids of cross-correlations over broad ranges in
temperature and rotational velocity (since these are the parameters
that affect the radial velocities the most), for an adopted surface
gravity of $\log g = 3.5$ (based on the system's probable pre--main
sequence status) and solar composition. The values obtained are
$T_{\rm eff} = 5900$~K and $v \sin i = 28$~\kms\ for the primary star,
and $T_{\rm eff} = 5800$~K and $v \sin i = 22$~\kms\ for the
secondary, with estimated uncertainties of $\sim$150~K and 1~\kms,
respectively. These temperatures are consistent with the reported
spectral type G0. We see no evidence in our spectra of the phenomenon
of veiling, which is common in many other young stars and was reported
by \cite{Arellano:03} to be present in HD~34700. However, veiling can
be variable. 

Table~\ref{tab:rvs} lists the radial velocities for both components,
referred to the heliocentric frame.  Typical uncertainties are given
below.  The stability of the zero-point of our velocity system was
monitored by means of exposures of the dusk and dawn sky, and small
systematic run-to-run corrections were applied in the manner described
by \cite{Latham:92}. The accuracy of the CfA velocity system, which is
within about 0.14~\kms\ of the reference frame defined by minor
planets in the solar system, is documented in the previous citation
and also by \cite{Stefanik:99} and \cite{Latham:02}. 

Following \cite{Zucker:94}, we have also determined the light ratio
between the secondary and the primary of HD~34700 at the mean
wavelength of our spectroscopic observations (5187~\AA), which is
close to the $V$ band: $l_B/l_A = 0.80 \pm 0.02$. \cite{Arellano:03}
measured the equivalent widths of 105 relatively unblended lines in
both components in one of their high resolution spectra, over the
wavelength range 4260--6770 \AA. The average ratio they found between
the lines of the secondary and those of the primary is $1.00 \pm
0.03$. Since the stars are of very similar temperature, the
line-strength ratio should be close to the light ratio between the
stars. The \cite{Arellano:03} value is somewhat higher than our own.
We adopt in the following the straight average, $l_2/l_1 = 0.9$. 

\begin{figure} 
\epsscale{1.0} 
\plotone{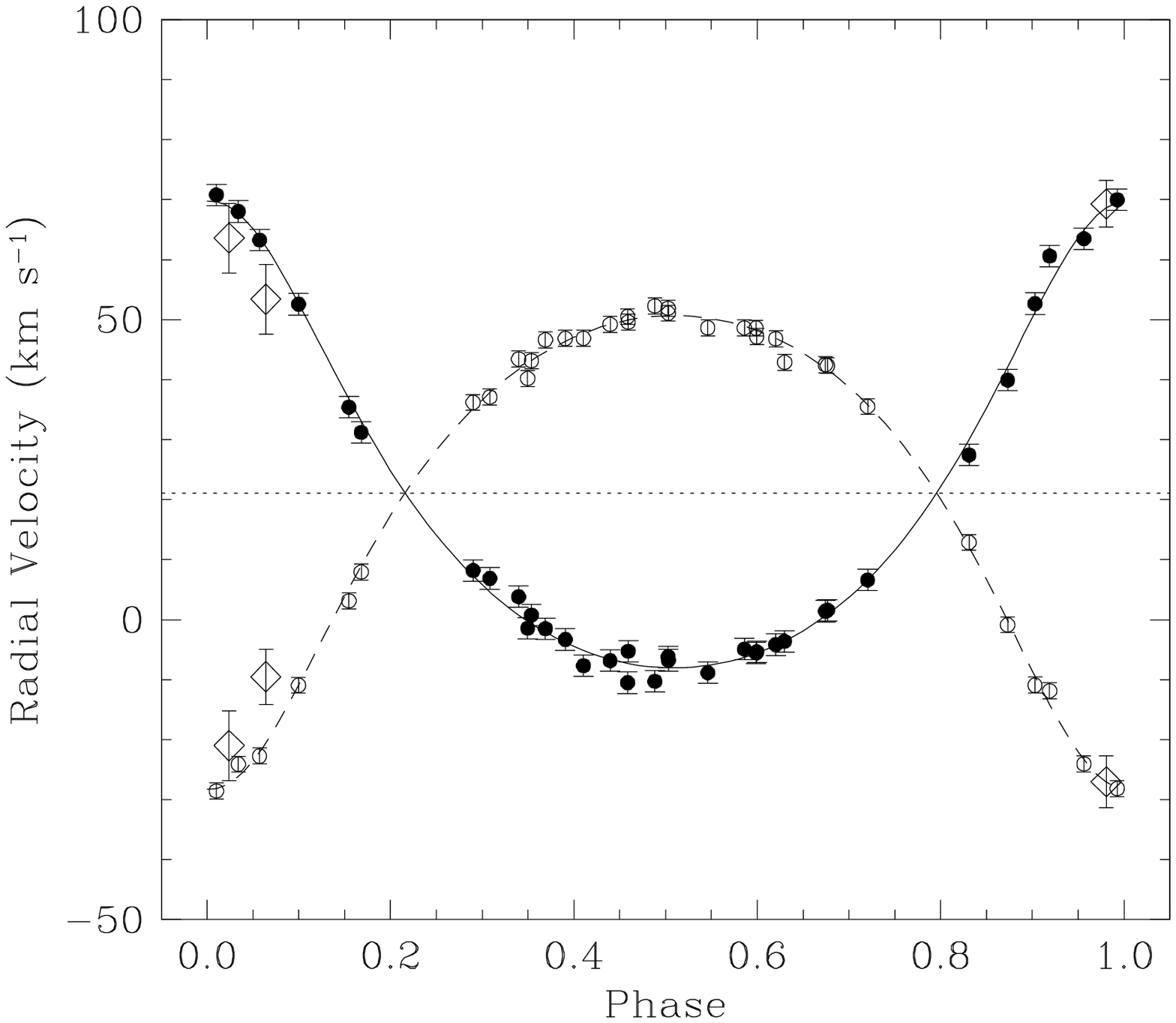}
 \figcaption[Torres.fig1.ps]{Double-lined orbital solution (solid
curve for primary, dashed curve for secondary) along with our radial
velocity measurements for HD~34000 (circles). Diamonds represent the
measurements by \cite{Arellano:03} (not used in the fit). Phase 0.0
corresponds to periastron passage, and the dotted line indicates the
systemic velocity.\label{fig:orbit}}
 \end{figure}

A double-lined orbital solution was easily obtained from our radial
velocities, showing a period of 23.5 days and a significant
eccentricity. This orbit is shown graphically in Fig.~\ref{fig:orbit},
and the elements are listed in Table~\ref{tab:elem}, where the symbols
have their usual meaning. RMS residuals for the primary and secondary
are 1.70~\kms\ and 1.34~\kms, respectively.  The slightly larger
errors for the primary are explained by its higher value of $v \sin
i$. Velocity residuals from our observations are listed in
Table~\ref{tab:rvs}. 

The heliocentric center-of-mass velocity we derive is $+21.03 \pm
0.18$~\kms. This is in excellent agreement with the radial velocity
for the CO material in the circumstellar disk measured by
\cite{Zuckerman:95}, which is $+$21~\kms\ (with a FWHM of 4.6~\kms\
for the $J = 1$--2 transition of CO from which the velocity was
measured).  We note also that this radial velocity is not far from
typical values in the Orion star-forming region, which are about
$+$25~\kms\ \citep[although with a spread of several \kms; see,
e.g.,][]{Hartmann:86}. 

In reporting double lines for HD~34700, \cite{Arellano:03} also
presented their radial velocity measurements for both components based
on the three high-resolution spectra they obtained. Those measurements
are in fair agreement with our orbit, as shown in
Fig.~\ref{fig:orbit}\footnote{Examination of the individual velocities
for each line in Table~2 of \cite{Arellano:03} leads us to believe
that the errors they quote in their Table~1 represent the scatter of
the measurements rather than the uncertainty of the mean (which should
be $\sqrt N$ smaller). If the smaller errors are adopted, then the
agreement with our orbit for two of their velocities is considerably
worse.}. 

\section{Discussion}
\label{sec:discussion}

As described in \S\ref{sec:intro}, there are fairly compelling
indications that HD~34700 is a young object from a number of
independent studies based on a wide variety of observational
techniques. \cite{Zuckerman:95} considered the age to be $\sim$10~Myr
or less, although based only on circumstantial evidence. The
fundamental difficulty is that the distance to HD~34700 is essentially
unknown, and therefore it cannot be placed on the H-R diagram and
compared to model isochrones in order to estimate the age. The star
was measured by the Hipparcos mission \citep{ESA:97}, but the
published parallax is rather uncertain ($\pi_{\rm HIP} = 0.86 \pm
1.84$~mas), and corresponds to a formal distance of 1160~pc.
Assumptions on the distance to HD~34700 in the literature have ranged
from 55~pc \citep{Sylvester:96b} to 90~pc \citep{Coulson:98}, derived
by adopting absolute magnitudes and intrinsic colors from the spectral
type, and accounting roughly for extinction. \cite{vandenAncker:98}
adopted a lower limit of 180~pc, and \cite{Bhatt:00} relied on the
Hipparcos determination. 

Based on the light ratio of $l_2/l_1 = 0.9$ (\S\ref{sec:orbit}) and
the apparent system magnitude of $V = 9.15$ \citep{ESA:97}, we infer
individual magnitudes of $V_1 = 9.85$ and $V_2 = 9.96$ assuming there
is no extinction. We may then use theoretical isochrones such as those
by \cite{Siess:97} \citep[see also][]{Siess:00} along with our
effective temperatures to compute the distance to each star assuming
they are located on the zero-age main sequence (ZAMS), and adopting
the solar value for the metallicity. This exercise results in
distances of 122~pc and 125~pc for the primary and secondary,
respectively. If the system were any closer, the stars would fall
below the ZAMS in the H-R diagram. An extinction value of $A_V = 0.2$
changes these minimum distances slightly to 111~pc and 114~pc. Thus we
conclude that the distance to HD~34700 is unlikely to be less than
about 100~pc, and therefore that the smaller values adopted to infer
properties of the system in most of the previous studies (which did
not account for the binarity of the object, since it was not known at
the time) are probably unrealistic. A parallax such as that
corresponding to 100~pc (10~mas) is large enough that Hipparcos would
most likely have been able to measure it accurately. A distance of
250~pc ($\pi = 4$~mas), on the other hand, is only about 2$\sigma$
from the nominal value reported in the catalog, and could have been
mismeasured.  At this distance the age inferred from the above models
would be about 9~Myr for both stars, in the absence of extinction.
HD~34700 is located in Orion ($\sim$8\arcdeg\ northwest of the belt),
though it is not particularly near any of the more conspicuous
star-forming regions in that area.  Perhaps the closest connection
that can be found is the similarity between its center-of-mass
velocity and the typical radial velocities of other young stars in
Orion, mentioned earlier\footnote{The proper motion components
reported both in the Hipparcos catalog and in the Tycho-2 catalog
\citep{Hog:00} are insignificantly small and are not of much help
regarding membership in Orion.}.  Nevertheless, if we assume the
star's distance to be the same as the main star-forming complexes, or
$\sim$460~pc, the age inferred from the brightness would be as young
as 3~Myr according to the model isochrones used above. 
	
Despite the quite extensive studies of HD~34700 over the past 15
years, it is rather surprising that a measurement of the Lithium
abundance ---one of the key indicators of youth-- has not been made
until very recently. \cite{Arellano:03} measured more than 100
spectral lines for both components of the binary, including
\ion{Li}{1}~$\lambda$6708. The results were not mentioned explicitly in
the text of their paper, but were reported in their Table~2 (available
only in electronic form). The Li equivalent width for the primary
(redward component in their spectra, according to our
Fig.~\ref{sec:orbit}) is 0.0881~\AA, and that for the secondary is
0.0790~\AA. Examination of their Figure~2 suggests, however, that the
Li line for the blueward component (secondary) may be blended with the
\ion{Fe}{1}~$\lambda$6705.1 line of the redward component, in which
case their Li equivalent width may be overestimated for the secondary.
We adopt the measurements at face value. Correction for binarity using
$l_2/l_1 = 0.9$ leads to final values of 0.17~\AA\ for both
components. 

Equivalent widths such as these are not large compared to typical
values in T~Tauri stars, but HD~34700 is considerably hotter than most
T~Tauri stars. Lithium depletion is a strong function of temperature,
and possibly also of rotational velocity. Comparison with diagrams of
Li equivalent width vs.\ temperature such as those by
\cite{Neuhauser:97} (their Fig.~3), \cite{Alcala:00} (Fig.~2), or
\cite{Wichmann:00} (Fig.~1), and others, indicate that the stars in
HD~34700 lie essentially on the upper envelope of the Pleiades
distribution (age $\sim$120~Myr), and that the Li strength is not
quite as strong as in stars in the younger cluster IC~2602 (age
$\sim$35~Myr), which have equivalent widths of up to about 0.25~\AA\
at these temperatures. We note, however, that if veiling were
significant, as discussed by \cite{Arellano:03}, the measured
equivalent widths for HD~34700 could be underestimated because the
lines may be filled in by featureless continuum radiation. Therefore,
the age of roughly 100~Myr or so is only an upper limit. Nevertheless,
on the basis of the Li strength alone it is quite possible that the
age is several tens of Myr instead of a few Myr, and therefore that it
is not as young as some of the other evidence presented above would
seem to suggest. As proposed by \cite{Fujii:02}, HD~34700 could be a
young G star near the end of the T~Tauri phase.  From Fig.~5 in
\cite{Alcala:00} and using the effective temperatures we determined
for each star, we estimate the abundance of Li as $\log N$(Li) $=$
3.1--3.2 on a scale where $\log N$(H) $= 12$. 

HD~34700 shows H$\alpha$ in emission, and the line profile changes
significantly on very short timescales ($\sim$1 day), as illustrated
by \cite{Arellano:03}. The equivalent width they measured is only
0.6~\AA, which places it in the weak-line T~Tauri class as opposed to
the classical T~Tauri group\footnote{A similar measure of the
H$\alpha$ emission ($\sim$0.2~\AA) can be derived from Fig.~2 of
\cite{Zuckerman:94}.}. Profile variations in other lines have also
been reported by \cite{Arellano:03}. Such changes are not uncommon in
young stars. 

The object is also a strong X-ray source, with a ratio of X-ray to
optical flux of $4.3 \times 10^{-4}$. It is listed in the ROSAT
All-Sky Survey Bright Source Catalogue \citep{Voges:99} under the
designation 1RXS~J051945.3$+$053509, and its X-ray properties (flux,
hardness ratios, etc.) are consistent with those of other young stars
\citep[see, e.g.,][]{Feigelson:99}. 

\begin{figure} 
\epsscale{1.0}
\vskip -1in
\plotone{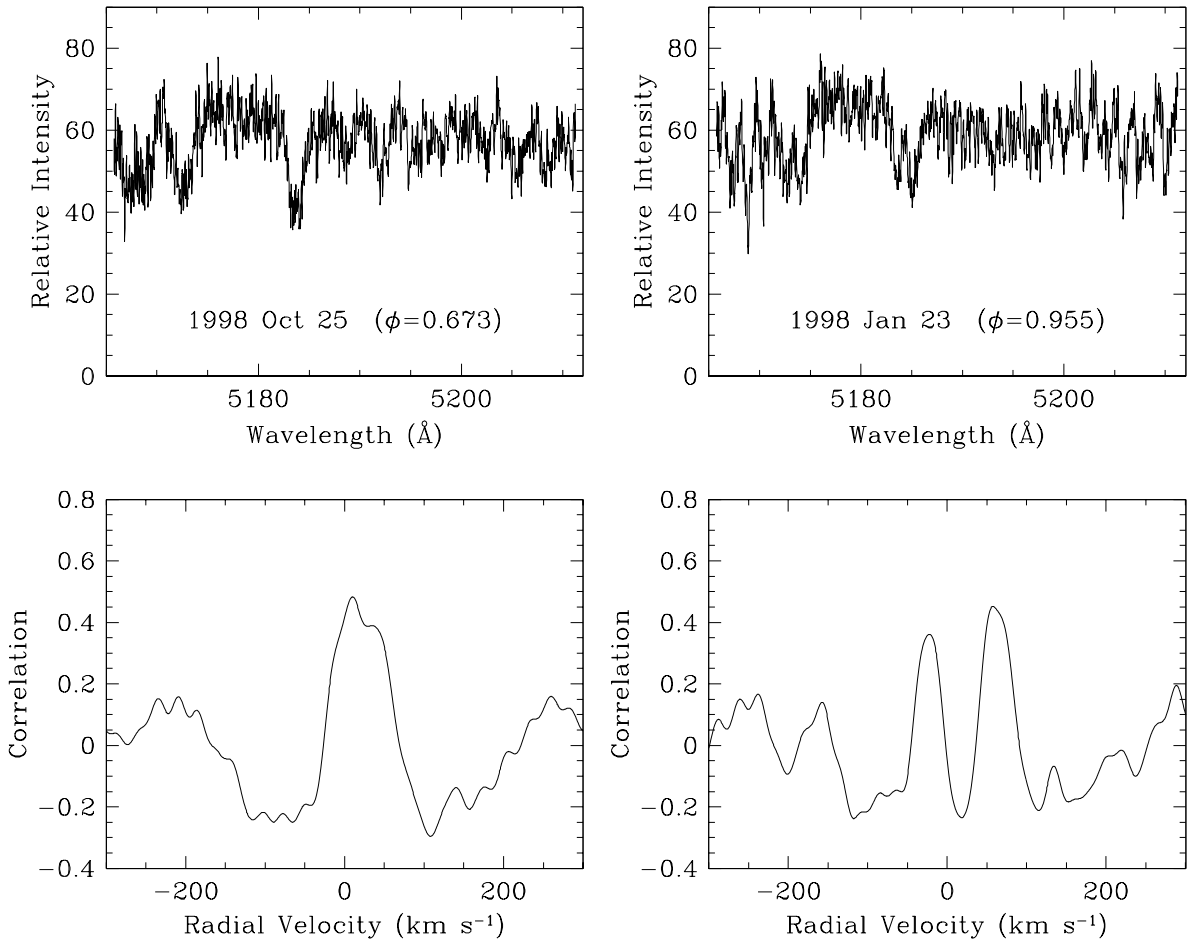}
 \figcaption[Torres.fig2.ps]{Left panels: Spectrum of HD~34700 and the
corresponding cross-correlation function for our observation of 1998
October 25, showing that the correlation peaks for the two components
are severely blended at this phase ($\phi = 0.673$; see
Fig.~\ref{fig:orbit}). Coincidentally this is the same date as one of
the observations by \cite{Mora:01}, on the basis of which they
reported a $v \sin i$ measurement of 46~\kms.  Right panels: Spectrum
and correlation function for 1998 January 23 ($\phi = 0.955$), showing
the correlation peaks well separated.\label{fig:badvsini}}
 \end{figure}

The projected rotational velocity measurements we reported earlier
(28~\kms\ and 22~\kms\ for the primary and secondary, respectively;
\S\ref{sec:obs}) are quite close to those given by \cite{Arellano:03}
(25~\kms\ and 23~\kms), although very different from the $v \sin i$ of
$46 \pm 3$~\kms\ determined by \cite{Mora:01}. The latter value is
based on two high-resolution spectra ($\lambda/\Delta\lambda =
49,\!000$, slightly higher than the resolution of our own
observations) taken at similar orbital phases of 0.67 and 0.72 (see
Fig.~\ref{fig:orbit}). By chance one of our exposures was obtained on
the same night as the first of the Mora et al.\ spectra.
Fig.~\ref{fig:badvsini} shows that the lines in that spectrum are
severely blended due in part to their intrinsic broadening, which
explains the large $v \sin i$ measurement by \cite{Mora:01}. The
velocity separation at this phase is $\sim$40~\kms. At other phases
the lines are clearly separated, as seen in the figure. 

The $v \sin i$ values for the components of HD~34700 are relatively
large, and one may ask whether they could be the result of tidal
locking of the rotation with the orbital motion of the binary, which
often goes together with orbital circularization in closer systems.
The significant eccentricity of the orbit and relatively long period
of 23.5 days seem to argue against this, although as a rule orbital
circularization occurs on much longer timescales than rotational
synchronization \citep[see, e.g.,][]{Hut:81}.  If the rotation period
$P_{\rm rot}$ of each star were the same as the orbital period, the
measured $v \sin i$ values and the relation
 \begin{equation} 
 v \sin i = {2\pi\over P_{\rm rot}} R \sin i
 \end{equation}
 lead to values for the projected radii of $R_1 \sin i =
13.0$~R$_{\sun}$ and $R_2 \sin i = 10.2$~R$_{\sun}$. These seem much
too large for PMS stars of this temperature. Even if the rotational
period corresponded to the orbital motion at periastron
(``pseudo-synchronization"), eq.(42) of \cite{Hut:81} predicts a
correction factor for $P_{\rm rot}$ of 0.72 at the measured
eccentricity of HD~34700. The projected radii would then be $R_1
\sin i = 9.4$~R$_{\sun}$ and $R_2 \sin i = 7.4$~R$_{\sun}$, still too
large in absolute terms by the time the increase due to the projection
factor is accounted for.  For comparison, at the distance of Orion
(and an age of 3~Myr) the same theoretical isochrones used above
predict sizes for the stars of about 4.4~R$_{\sun}$. Thus, the
components in this system are spinning much more rapidly than
synchronization would imply (at least twice as fast).  This may
perhaps be considered one more indication of youth, since main
sequence stars of this spectral type do not usually rotate at more
than a few \kms\ \citep[e.g.,][]{Gray:92}.

Based on the PMS models by \cite{Siess:97} and assuming that HD~34700
is at a representative distance of 200~pc, we estimate that the
absolute masses of the components are roughly 1.1--1.2~M$_{\sun}$,
similar to what is expected for main sequence stars of the same
spectral type.  The minimum masses from our orbital solution
(Table~\ref{tab:elem}), in turn, imply an inclination angle for the
orbit of about 50\arcdeg. The semimajor axis is then 0.21~AU, which is
well within the inner radius of the circumstellar disk as modeled by
\cite{Sylvester:96a} and \cite{Sylvester:96b}. At the same distance of
200~pc the angular semimajor axis would be 1~mas, difficult to resolve
with current interferometers, although the eccentricity of the orbit
would make the actual separation some 25\% larger at certain phases.
If the system is at the distance of Orion, the masses inferred from
the evolutionary tracks are about twice as large (and the orbital
inclination angle $\sim$39\arcdeg), the linear semimajor axis only
slightly larger than before (0.26~AU), and the angular semimajor axis
would be about 0.6~mas. Detecting the motion of the \emph{photocenter}
of the pair is far beyond the capabilities of the Hipparcos mission
(initially we had considered the unmodeled photocentric motion as a
possible explanation for the large error in the parallax).  From the
mass ratio and light ratio determined here, we estimate the semimajor
axis of the photocenter to be only 2.3\% of the angular semimajor axis
of the relative orbit. This is mostly because the stars are of similar
brightness, so their center of light does not move much.  At the
distance of Orion, the 14~$\mu$as signal would be challenging to
measure even for NASA's Space Interferometry Mission. 

Finally, we note that the Hipparcos epoch photometry ($H_p$ band)
shows no sign of photometric variability at the few milli-magnitude
level over the duration of the satellite's 3-year mission, which is
somewhat unexpected if the star is very young, but perhaps not so much
so if it is already near the ZAMS. Similar evidence in the $JHK$ bands
was presented by \cite{Eiroa:01}. 

\section{Concluding remarks}

We have confirmed the young star HD~34700 to be a double-lined
spectroscopic binary, and presented an accurate orbital solution with
a period of 23.4877~days and a significant eccentricity ($e = 0.2501
\pm 0.0068$). The components are of very nearly equal mass,
temperature, and luminosity. The measured projected rotational
velocities indicate super-synchronous rotation in both stars. 

We have examined the available indicators of youth, and find that when
considered together, the following present fairly compelling evidence
in favor of PMS status:

\noindent (a) Significant infrared excess suggesting the presence of a
sizeable circumstellar disk. Polarization and CO measurements
consistent with this idea. Large fractional luminosity in the
infrared. 

\noindent (b) H$\alpha$ in emission, with a strength ($\sim$0.6~\AA\
equivalent width) that places the object in the weak-line T~Tauri
class. 

\noindent (c) Strong \ion{Li}{1}~$\lambda$6708 absorption (equivalent
width 0.17~\AA). 

\noindent (d) Variable line profiles for H$\alpha$, and possibly other
spectral lines. Possibility of veiling. 

\noindent (e) Strong X-ray emission as seen by the ROSAT satellite,
consistent with that detected in other weak-line T~Tauri stars. 

\noindent (f) Rapid rotation of the components (28~\kms\ and 22~\kms,
$\sim$10 times faster than main-sequence stars of similar spectral
type) in an orbit that is wide enough that the stars are not tidally
synchronized. 

\noindent (g) Location of the star in the general vicinity of the
Orion star-forming region, and a radial velocity ($+21$~\kms) fairly
close to typical values for other young stars in that area. 

It is difficult to estimate a precise age for the system due to the
lack of an accurate parallax. If it is at the distance of Orion,
theoretical isochrones indicate it is only a few Myr old. However, it
could be closer and therefore be approaching the ZAMS, in which case
the age could be several tens of Myr. The \ion{Li}{1}~$\lambda$6708
absorption is not quite as strong as seen in other very young stars,
but the Li measurements for HD~34700 may perhaps be affected by
veiling. 

HD~34700 thus joins the relatively reduced group of a few dozen young
spectroscopic binaries with known orbits, with its double-lined nature
making it a particularly interesting case. 
	
\acknowledgements 

This research has made use of the SIMBAD database, operated at CDS,
Strasbourg, France, and of NASA's Astrophysics Data System Abstract
Service. We are grateful to A.\ Arellano-Ferro for providing the
electronic version of their Table~2 in advance of publication, to R.\
Neuh\"auser for pointers to the X-ray activity, and to the referee for
helpful comments. Partial support for this work from NASA's MASSIF SIM
Key project (JPL grant 1240033) is also acknowledged.

\clearpage

\clearpage

\clearpage

\begin{deluxetable}{crrrrc}
\tablenum{1}
\tabletypesize{\scriptsize}
\tablewidth{0pc}
\tablecaption{Radial velocity measurements of HD~34700 (heliocentric frame).\label{tab:rvs}}
\tablehead{
\colhead{HJD} & \colhead{RV$_{\rm 1}$} & \colhead{RV$_{\rm 2}$} &
\colhead{(O$-$C)$_{\rm 1}$}  & \colhead{(O$-$C)$_{\rm 2}$} & \colhead{Orbital} \\
\colhead{\hbox{~~(2,400,000$+$)~~}} & \colhead{(\kms)} &
\colhead{(\kms)} & \colhead{(\kms)} & \colhead{(\kms)} & \colhead{Phase\tablenotemark{a}} }
\startdata
    50154.6196\dotfill &  $+$59.90 &  $-$11.87 &   $+$4.30 &   $+$2.11  &   0.918 \\
    50382.7060\dotfill &   $-$3.56 &  $+$42.73 &   $+$0.06 &   $-$3.27  &   0.629 \\
    50411.9077\dotfill &  $+$39.65 &   $-$0.94 &   $-$2.34 &   $-$0.74  &   0.872 \\
    50448.7009\dotfill &   $-$6.86 &  $+$49.23 &   $-$0.20 &   $+$0.15  &   0.439 \\
    50710.8038\dotfill &   $-$5.50 &  $+$48.42 &   $+$0.17 &   $+$0.35  &   0.598 \\
    50729.8705\dotfill &   $-$7.53 &  $+$46.76 &   $-$2.33 &   $-$0.83  &   0.410 \\
    50752.8929\dotfill &   $-$3.27 &  $+$46.66 &   $+$0.61 &   $+$0.39  &   0.390 \\
    50781.7730\dotfill &   $-$4.28 &  $+$46.47 &   $+$0.04 &   $-$0.24  &   0.620 \\
    50804.7755\dotfill &   $-$5.30 &  $+$46.88 &   $+$0.32 &   $-$1.14  &   0.599 \\
    50836.6429\dotfill &  $+$63.33 &  $-$23.95 &   $-$1.40 &   $-$0.72  &   0.956 \\
    50845.6472\dotfill &   $+$3.74 &  $+$43.21 &   $+$3.00 &   $+$1.63  &   0.339 \\
    50854.5896\dotfill &   $+$6.32 &  $+$35.56 &   $-$0.28 &   $-$0.09  &   0.720 \\
    50872.6278\dotfill &  $-$10.28 &  $+$52.14 &   $-$2.31 &   $+$1.74  &   0.488 \\
    50900.5368\dotfill &   $+$1.28 &  $+$42.12 &   $+$0.47 &   $+$0.61  &   0.676 \\
    51061.8857\dotfill &   $-$8.74 &  $+$48.45 &   $-$1.06 &   $-$1.67  &   0.546 \\
    51080.8535\dotfill &   $+$0.67 &  $+$43.00 &   $+$1.39 &   $-$0.06  &   0.353 \\
    51102.8486\dotfill &   $+$8.04 &  $+$36.35 &   $+$0.86 &   $+$1.29  &   0.290 \\
    51111.8753\dotfill &   $+$1.39 &  $+$42.33 &   $+$0.83 &   $+$0.57  &   0.674 \\
    51126.7643\dotfill &   $+$6.68 &  $+$37.16 &   $+$2.11 &   $-$0.55  &   0.308 \\
    51154.8194\dotfill &   $-$6.11 &  $+$51.72 &   $+$1.97 &   $+$1.20  &   0.502 \\
    51174.7082\dotfill &   $-$1.56 &  $+$40.03 &   $-$1.25 &   $-$2.62  &   0.349 \\
    51198.6539\dotfill &   $-$1.54 &  $+$46.46 &   $+$0.62 &   $+$1.94  &   0.369 \\
    51240.5926\dotfill &  $+$35.28 &   $+$2.74 &   $-$1.44 &   $-$2.40  &   0.154 \\
    51467.8701\dotfill &  $+$28.09 &  $+$12.33 &   $-$1.86 &   $+$0.34  &   0.831 \\
    51522.7615\dotfill &  $+$31.40 &   $+$7.55 &   $-$1.58 &   $-$1.37  &   0.168 \\
    51542.5413\dotfill &  $+$70.80 &  $-$28.28 &   $+$1.34 &   $-$0.26  &   0.010 \\
    51544.6439\dotfill &  $+$52.37 &  $-$10.82 &   $-$0.28 &   $+$0.18  &   0.099 \\
    51566.5918\dotfill &  $+$67.81 &  $-$24.01 &   $+$0.33 &   $+$2.01  &   0.034 \\
    51576.5787\dotfill &   $-$5.35 &  $+$49.46 &   $+$2.02 &   $-$0.34  &   0.459 \\
    51577.6052\dotfill &   $-$6.70 &  $+$51.04 &   $+$1.38 &   $+$0.52  &   0.503 \\
    51590.6213\dotfill &  $+$63.32 &  $-$22.72 &   $-$0.16 &   $-$0.75  &   0.057 \\
    51610.4871\dotfill &  $+$52.42 &  $-$10.79 &   $+$1.36 &   $-$1.40  &   0.903 \\
    51612.5918\dotfill &  $+$69.90 &  $-$27.96 &   $+$0.61 &   $-$0.10  &   0.992 \\
    51623.5376\dotfill &  $-$10.50 &  $+$50.26 &   $-$3.15 &   $+$0.48  &   0.458 \\
    51626.5283\dotfill &   $-$5.03 &  $+$48.35 &   $+$1.28 &   $-$0.37  &   0.586 \\
\enddata
\tablenotetext{a}{Referred to the time of periastron passage.}
\end{deluxetable}

\clearpage

\begin{deluxetable}{lccc@{}cc}
\tablenum{2}
\tablecolumns{2}
\tablewidth{25pc}
\tablecaption{Spectroscopic orbital solution for HD~34700.\label{tab:elem}}
\tablehead{
\colhead{\hfil~~~~~~~~~~~~~~Parameter~~~~~~~~~~~~~~~} & \colhead{Value}}
\startdata
\multicolumn{2}{l}{Adjusted quantities} \\
~~~~$P$ (days)\dotfill                       &  23.4877~$\pm$~0.0013\phn     \\
~~~~$\gamma$ (\kms)\dotfill                  &  $+21.03$~$\pm$~0.18\phs\phn       \\
~~~~$K_{\rm 1}$ (\kms)\dotfill               &  38.83~$\pm$~0.42\phn         \\
~~~~$K_{\rm 2}$ (\kms)\dotfill               &  39.33~$\pm$~0.34\phn           \\
~~~~$e$\dotfill                              &  0.2501~$\pm$~0.0068           \\
~~~~$\omega_1$ (deg)\dotfill                 &  358.1~$\pm$~1.6\phn\phn \\
~~~~$T$ (HJD$-$2,400,000)\tablenotemark{a}\dotfill    &  51072.558~$\pm$~0.093\phm{2222} \\
\multicolumn{2}{l}{Derived quantities} \\
~~~~$M_{\rm 1}\sin^3 i$ (M$_{\sun}$)\dotfill &  0.531~$\pm$~0.011         \\ 
~~~~$M_{\rm 2}\sin^3 i$ (M$_{\sun}$)\dotfill &  0.524~$\pm$~0.012       \\
~~~~$q\equiv M_{\rm 2}/M_{\rm 1}$\dotfill    &  0.987~$\pm$~0.014       \\
~~~~$a_{\rm 1}\sin i$ (10$^6$ km)\dotfill    &  12.14~$\pm$~0.13\phn    \\
~~~~$a_{\rm 2}\sin i$ (10$^6$ km)\dotfill    &  12.30~$\pm$~0.11\phn      \\
~~~~$a \sin i$ (R$_{\sun}$)\dotfill          &  35.12~$\pm$~0.25\phn      \\
\multicolumn{2}{l}{Other quantities pertaining to the fit} \\
~~~~$N_{\rm obs}$\dotfill                    &  35                      \\
~~~~Time span (days)\dotfill                 &  1472                  \\
~~~~$\sigma_{\rm 1}$ (\kms)\dotfill          &  1.70      \\
~~~~$\sigma_{\rm 2}$ (\kms)\dotfill          &  1.34      \\
\enddata
\tablenotetext{a}{Time of periastron passage.}
\end{deluxetable}

\end{document}